# High spatial resolution diffraction diagnostics for intense terahertz sources

Shang-qing Li,[1,2] Jing-long Ma,[1] Xiao-jun Wu,[3,*] Bao-long Zhang,[1,2] Chen Ou-yang,[1,2] Tian-ze Wang,[1,2] Dan Wang,[1,2] Xuan Wang,[1] and Yu-tong Li[1,2,4,*]

[1]Beijing National Laboratory for Condensed Matter Physics, Institute of Physics, Chinese Academy of Sciences, Beijing 100190, China
[2]School of Physical Sciences, University of Chinese Academy of Sciences, Beijing 100049, China
[3]School of Electronic and Information Engineering, Beihang University, Beijing, 100191, China
[4]Songshan Lake Materials Laboratory, Dongguan, Guangdong 523808, China
*Corresponding author: xiaojunwu@buaa.edu.cn; ytli@iphy.ac.cn

**We propose a high resolution spatial diagnostic method via inserting a millimeter-gap grating into the collimated terahertz beam to monitor the minute variation of the terahertz beam in strong-field terahertz sources, which is difficult to be resolved in conventional terahertz imaging systems. To verify the method, we intentionally fabricate tiny variations of the terahertz beam through tuning the iris for the infrared pumping beam before the tilted-pulse-front pumping (TPFP) setups. The phenomena can be well explained by the the theory based on tilted pulse front technique and terahertz diffraction. We believe our observation not only help further understand the mechanism of intense terahertz generation, but also may be useful for strong-field terahertz applications.**

http://dx.doi.org/10.1364/OL.99.099999

Compatibility of the Ti:sapphire and Ytterbium high energy femtosecond laser technologies with lithium niobate crystal-based tilted pulse front technique has made the strong-field terahertz sources ubiquitous [1-16]. For intense terahertz radiation process, it is possible to deduce the emission mechanism through systematically characterizing the generated terahertz properties including spectrum, polarization, and beam profiles [17]. In real applications, it is also highly demanded to monitor the terahertz beam profile and its positions for better focusing resulting in strong fields, because most observations related to field-induced phenomena are based on small modulation responses of the materials or substances pumped by strong-field terahertz waves [18]. In weak-field terahertz science and technology, it is extremely difficult to direct image terahertz beam profiles. Until recently, intense terahertz sources and commercial terahertz pyroelectric cameras become available which provides the feasibility to directly investigate or monitor the terahertz beam profile. However, due to the long wavelength of terahertz waves, the focused terahertz beam size is around millimeter-level. Small variation of the terahertz beam is difficult to be detected. Previously, the terahertz beam size and position variation in the generation process of tilted pulse front technique in lithium niobate has been systematically investigated by combining imaging magnification method and terahertz cameras [17]. With this method, millimeter level change can be well resolved.

In our work, we propose a grating diffraction method to diagnose the tiny variation of the generated terahertz beam from lithium niobate crystal driven by Ti:sapphire laser pulses. Furthermore, we experimentally fabricate small changes of the radiated terahertz beam via inserting an iris before the IR pumping beam entering into the TPFP setup. The changes are detected by the proposed method and the observations can be well explained by the theory of tilted pulse front technique and terahertz diffraction.

To realize the aforementioned diagnostic function, we implement the following experiments. The experimental setup is illustrated in Fig. 1(a). An intense terahertz beam source radiated from lithium niobate crystal driven by a commercial Ti:sapphire laser amplifier (Amplitude System, Pulsar 20, central wavelength 800 nm, pulse duration 30 fs, repetition rate 10 Hz) via tilted pulse front technique is generated. Fig. 1(b) and (c) plots the terahertz temporal waveform and its corresponding Fourier transform spectrum extracted from the stroboscopically electro-optic sampling in 100 μm thick ZnTe crystal. The single-cycle terahertz pulse is ~8 ps and its central frequency is ~0.3 THz with 0-1 THz frequency range. As is shown in Fig. 1(a), the emitted terahertz waves are collimated and focused by two parabolic mirrors into a terahertz camera (Spiricon, Pyrocam IV). Under 50 mJ single pulse energy pumping, the generated terahertz energy is ~50 μJ. The focused terahertz spot profile is shown in Fig. 1(d). A grating (named as terahertz grating hereafter) with two gaps of 15 mm width, 30 mm period, and 0.2 mm thickness is inserted into the collimated terahertz beam to simulate the diffraction effect, as shown in inset of Fig. 1(a). The present of terahertz grating would transform the image received by terahertz camera from Airy spots to diffraction patterns, and the position of terahertz camera is needed to be optimized (L = 150 mm) to obtain tight images.

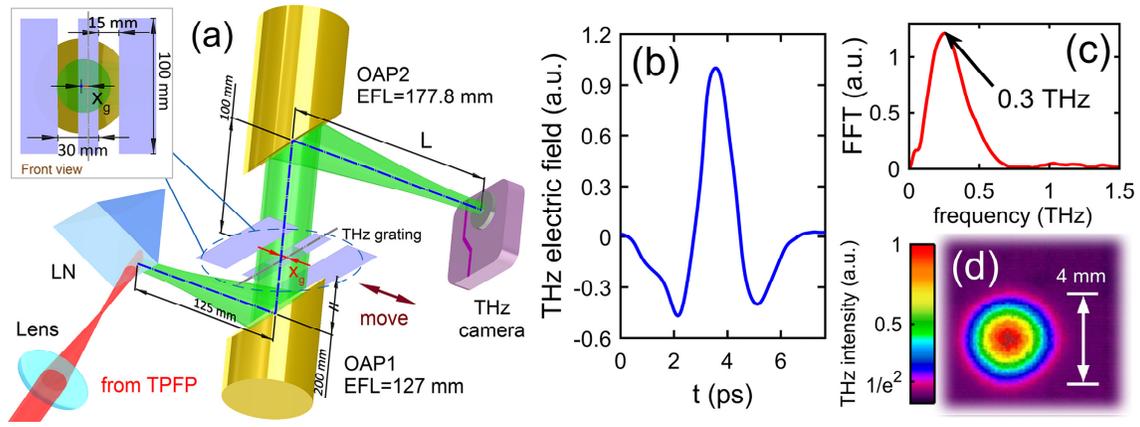

Fig. 1. High spatial resolution diffraction diagnostic system and the matched intense terahertz source. (a) Schematic diagram of the experimental setup for the terahertz diffraction method. (b) Typical terahertz temporal waveform, and (c) its corresponding spectrum. (d) Focused terahertz beam profile recorded by a terahertz camera without the terahertz grating when L = 177.8 mm.

With the aforementioned experimentally obtained diffraction diagnostics system and the intense terahertz source, we ought to analyze the modulation to diffraction patterns while inserting the terahertz grating for the first step. A two-dimensional (2D) finite-difference time-domain (FDTD) simulation using the commercial software package (Lumerical FDTD, Lumerical Solutions Inc.) is carried out. In order to reduce computational load, we shrink the length of collimated terahertz beam before terahertz grating (from 20 cm to 5 cm) in our simulation, and the change would not impact the authenticity of the simulation results. The calculated diffraction patterns are produced and shown in Fig. 2(a)-(d). In this case, the terahertz source position is fixed but the grating is movable. When the centers for both of the grating and the terahertz beam are perfectly overlapped, a bright beam spot accompanied by two weak diffraction patterns are symmetrically distributed on the left and right sides of the bright spot (see the inset of Fig. 2(b)). When the grating is deviated from the beam center, symmetry breaking happens and asymmetrical diffraction patterns are formed (see Fig. 2(a) and (c)). The experimental results shown in the insets of Fig. 2(a)-(c) well reproduce the theoretical prediction. Thus, the asymmetry of the diffraction pattern could reveal the deviation of terahertz beam and terahertz grating qualitatively. Technically and quantitatively, we need to develop the inversion method like Yang-Gu algorithm to deduce the terahertz beam profile on terahertz grating from the diffraction patterns gained by terahertz camera. However such method is difficult to achieve and we propose an alternative and uncomplicated way to portray the modulation aforementioned. In our simulation and experiments, we find that there is an apparent linear relationship between the terahertz grating position $x_g$ and the peak intensity position $x_p$ in diffraction patterns as shown in Fig. 2(d). The fitted slope can be estimated as

$$\beta \approx <x_p/x_g> \approx 0.075. \qquad (1)$$

It is an empirical correlation and we could apply the formula to derive the geometric relationship between the collimated terahertz beam and terahertz grating facing a diffraction pattern gained by terahertz camera. Moreover, since the transition motion for the grating and the generated terahertz beam are relative, the diffraction method can be applied to diagnostic the minute movement of the terahertz beam when terahertz grating position is fixed. We can examine the idea for the next step.

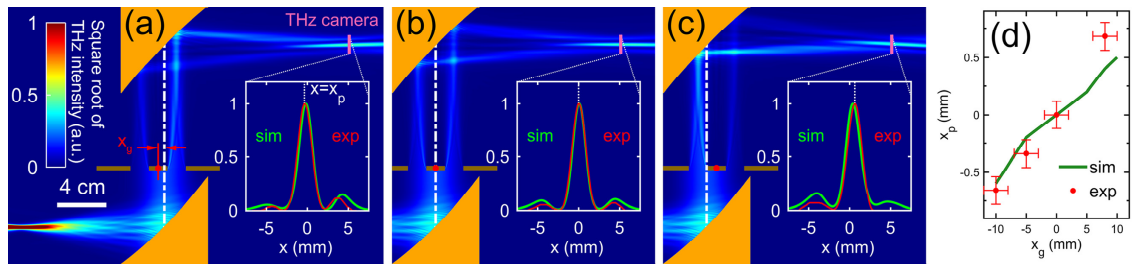

Fig. 2. Experimental and simulation results illustrating the relevance between terahertz diffraction pattern received by terahertz camera and terahertz grating position. (a)-(c) Meridional diffraction fields and the corresponding diffraction patterns for different grating positions x=-5 mm, 0, and 8 mm respectively. (d) Peak intensity position $x_p$ in terahertz diffraction patterns as a function of terahertz grating position $x_g$.

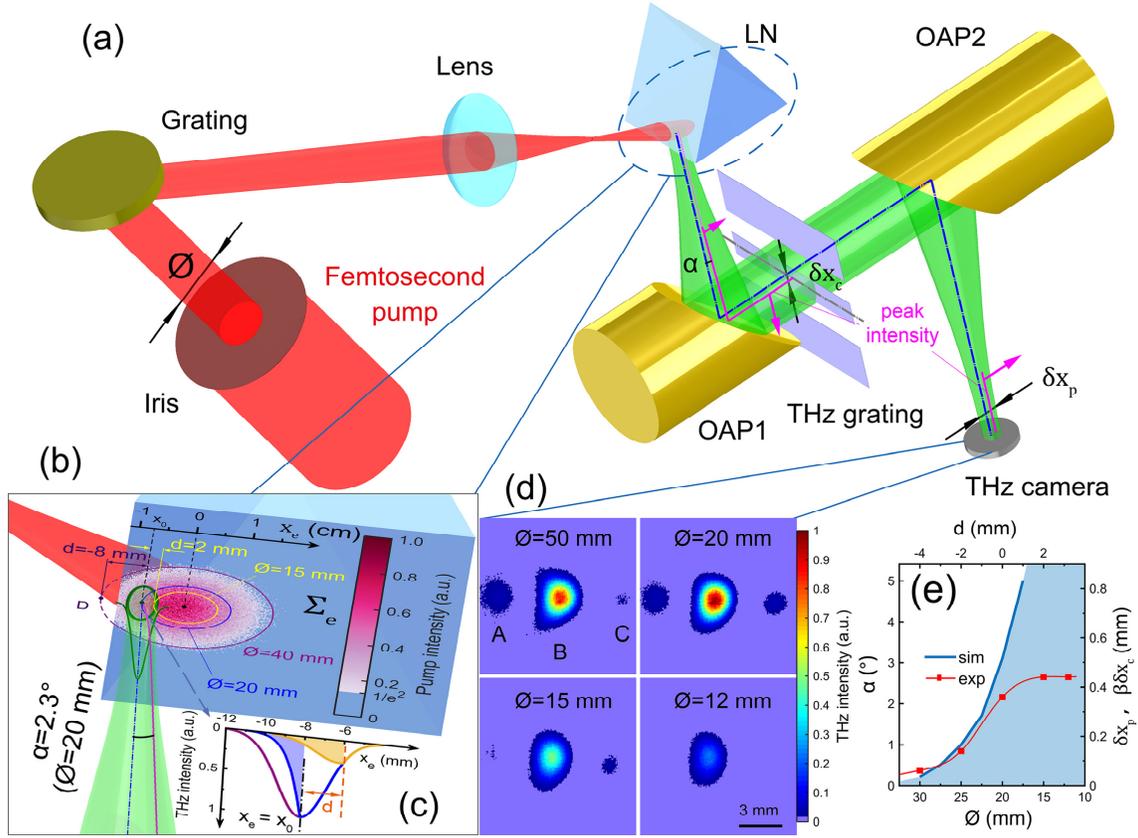

Fig. 3. Diffraction diagnostic for monitoring the moving of the terahertz beam. (a) Experimental setup to diagnose a tiny variation of the terahertz beam direction and position. (b) Pump shrinks and terahertz emits atilt when closing the iris. (c) Calculated terahertz beam profile in lithium niobate crystal for different iris size. Dark purple line corresponds to Ø=40 mm, and blue, 20 mm, yellow, 15 mm, respectively. (d) Experimental diffraction patterns received by terahertz camera. (e) Extracted experimental movement of the terahertz beam compared with the theoretical calculations.

To verify the feasibility and the spatial resolution sensitivity of the proposed diffraction method, we intentionally design a minute movement of the terahertz beam by inserting an iris with variable diameter into the pumping beam before the tilted pulse fronts setup, as shown in Fig. 3(a). The infrared pumping beam diameter is ~40 mm ($1/e^2$). The maximum iris diameter is 50 mm. After the modulation by the inserted iris, it illuminates onto a grating with a line density of 1500 lines/mm. The diffracted beam propagates through a lens (focal length=85 mm) and illuminates the lithium niobite crystal. Owing to optical rectification (OR) the intense terahertz is excited by the tilted-front pump and emit from rear surface (labeled as $\Sigma_e$). The terahertz wave is received by the diffraction diagnostics system in Fig. 1(a). In order to characterize the association between diffraction patterns and the movement of terahertz beam more clearly and intuitively, we shift the terahertz grating position ($x_g = 5$ mm), as is shown in Fig. 3(a), and the position of terahertz grating is locked hereafter. Fig. 3(d) shows the experimental results of the grating diffraction diagnostic when varying the iris diameter. When the iris diameter is fully open with 50 mm diameter, the diffraction pattern is asymmetric with a brighter beam spot on the left side (A) of the principal beam spot

(B). When the iris diameter decreases to 20 mm, the diffraction pattern tends to be symmetric with two uniform and bright spots around the main maximum. Further reducing the iris diameter to 15 mm, the spot A disappears while the spot C is still obvious. Both spot A and C disappear when the iris diameter is smaller than 12 mm. Comparing with diffraction patterns in Fig. 2 and Fig. 3(b), it is believable that terahertz beam spot changes its location on terahertz grating when varying the iris. The concrete direction of terahertz beam is indicated in Fig. 3(a) by magenta lines when Ø = 20 mm, and the magenta arrows imply the movement of terahertz beam direction when Ø is decreasing.

To explain these phenomena we develop the unilaterally blocked Gauss beam model illustrated in Fig. 3(b-c). According to the previous experiments in our team [19, 20], the pump beam center and the terahertz beam center are not coincident, and the distance between them is 8.0±2.0 mm. When closing the iris the terahertz spot profile would be reshaped on surface $\Sigma_e$ due to absence of OR, as shown in Fig. 3(c). Distribution of the reshaped terahertz spot profile is expressed as:

$$I(x_e) = \begin{cases} I_1(x_e) \sim \exp(-2(x_e - x_0)^2 / r^2) & (x_e \geq x_0 + d) \\ I_2(x_e) \leq I_1(x_e)\exp(-\dfrac{\alpha_{THz}}{\tan\gamma}(x_0 + d - x_e)) & (x_e < x_0 + d) \end{cases},$$

where $\alpha_{THz} = 17$ cm$^{-1}$ and $\gamma = 62°$ are the intensity absorption coefficient and phase matching angle of lithium niobite crystal respectively. $r = 2.5$ mm is the horizontal radius (1/e$^2$) of terahertz spot [19]. Other parameters are defined in Fig. 3(b). The reshaped terahertz Gauss beam with deficiency would no longer propagate along the direction perpendicular to the surface $\Sigma_e$, and it would emit a tilt as shown in Fig. 3(b). To be more precise, we conduct series 2D FDTD simulations to model the propagation of the reshaped terahertz beam and exploit it to resolve $\alpha$-$d$ unequal relation (indicated by the blue curve in Fig. 3(e) on condition of $I_2(x_e) = 0$. Thus the experimental $\alpha$-$d$ curve should lies within the region marked by the nattier blue). meanwhile we can extract $\emptyset$-$\delta x_p$ relation from the experimental diffraction patterns (indicated by the red curve in Fig. 3(e). $\delta x_p$ is the movement of peak intensity on diffraction patterns). The two curves can be depicted in one graph because their coordinate are correlated. when $\alpha$ satisfies small-angle approximation, the approximate expressions are listed as below:

$$\alpha = \delta x_c / f_1 = \delta x_p / \beta f_1, \quad (2)$$

$$d = -C\emptyset - x_0, \quad (3)$$

where $f_1$ is the EFL of OAP1, $\delta x_c$ is the movement of peak intensity on the plane of terahertz grating, $C \approx 0.40$ is the linearity coefficient between $\emptyset$ and $d$. As is shown in Fig. 3(e), The experimental data are within theoretical predictions and the detected maximum varieties of feature parameters $\alpha$ (~3°) and $\delta x_c$ (~5 mm) are very close to the limit resolutions listed in Table 1. This paragraph exhibits an application of the proposed diffraction diagnostic method to explore the physics of the intense terahertz sources generated by TPFP and inspect the spatial resolutions.

Table 1. Resolutions of feature parameters of the proposed diffraction diagnostic method.

| parameters | resolutions | descriptions |
| --- | --- | --- |
| $\delta x_p$ | 80 μm | Resolution of THz camera* |
| $\delta x_c$ | 1.1 mm | Calculated by Eq. (1) |
| $\alpha$ | 0.50° | Calculated by Eq. (2) |

*defined by the distance between two adjacent pixels.

Furthermore, with this method, we can also estimate the terahertz central frequency. According to the Young's double-slit interreference experiments, the first order diffraction angle $\theta_1 = x_1/L \approx 4.5$ mm/150 mm = 1.7°, where $x_1$ is the first order diffraction pattern position, and the central wavelength $\lambda = \theta_1 d_g$ =1.7°×30 mm = 0.9 mm, $d_g$ is the period of terahertz grating. The calculated central frequency is ~0.3 THz, which agrees well with the values obtained in electro-optic sampling method (see Fig. 1(c)).

In summary, we introduce an effective diffraction method to diagnose the movement of the terahertz beam position and direction generated from lithium niobite crystal via tilted pulse fronts technique. The proposed method in theory could detect the minute revolvement of emitted angle (~0.5°) and displacement (~1.1 mm) of divergent and collimated terahertz beam respectively. To verify these spatial resolutions we experimentally corroborate the proposed approach through on purpose fabricate a ~3° revolvement of emitted angle of terahertz beam via inserting an iris into the infrared pumping beam before it enters into the tilted pulse fronts setup. The movement of terahertz beam is detected by the diffraction diagnostic system and well explained by the unilaterally blocked Gauss beam model. We believe our observation not only helps deeply understand the terahertz radiation physics, but may also have potential contributions for terahertz beam focusing, manipulation, and applications.

**Funding**. This work is supported by the Science Challenge Project (No. TZ2016005), National Nature Science Foundation of China (Grants Nos. 11827807, 61905007, 11520101003, 11861121001) and the Strategic Priority Research Program of the Chinese Academy of Sciences (Grant No. XDB16010200).

**Disclosures.** The authors declare no conflicts of interest.